\documentclass[aps,prb,twocolumn,showpacs]{revtex4-1}
\usepackage[dvips]{graphicx}
\usepackage{bm,amsmath}

\begin{document}

\title{Hot-electron Effect in A Cylindrical Nanoshell}

\author{Ya-Ni Zhao, Lin Zhang and Shi-Xian Qu\footnote{author to whom
correspondence should be addressed\\
e-mai: sxqu@snnu.edu.cn}} \affiliation{Institute of Theoretical \&
Computational Physics, School of Physics and Information Technology,
Shaanxi Normal University, Xi'an 710062, China}

\author{Michael R. Geller}
\affiliation{Department of Physics and Astronomy, University of
Georgia, Athens, Georgia 30602-2451}

\begin{abstract}
A general expression for calculating the thermal power transferring
from 3-dimensional electron to any $D$-dimensional phonon subsystem
is derived in this paper. Electron-phonon coupling in a free
suspended cylindrical nanoshell is studied, in which acoustic
phonons are confined to quasi-one dimension but electrons behave
three-dimensionally. The temperature dependence of the thermal power
is obtained analytically, and the low-temperature crossover from the
$T^3$ to $T^3/(1-\nu^2)+9\gamma T^4/[T^*(1-\nu^2)^{3/2}]$ dependence
is also observed. The corresponding quantities are estimated for the
material parameters from a metallic nanotube.
\end{abstract}

\pacs{63.22.+m, 63.20.Kr, 85.85.+j}

\maketitle
\clearpage

\section{Introduction}

The development of modern micro-manufacturing and extremely low
temperature technology enables people to explore the physical
properties of condensed matter in nanoscales, where the phonon
thermal conductance at low-temperature is one of the most active
area that attracts much interest of
researchers.\cite{ClelandBook,Schwab
etal,Roukes1999,Cleland2001,Blencowe2004} In addition to its
significance for investigating fundamental phonon physics and
macroscopic quantum phenomena, it may find applications in nanoscale
bolometers and calorimeters with unprecedented
sensitivity,\cite{Roukes1999,Cleland2002,Schmidt2004} and even in
quantum information processing.\cite{Armour,Zou}
%The coupling between electrons and phonons plays an important role in
%determining the thermal properties of these nanostructures.
It is an established fact that the conduction electrons in a metal
at low temperature are only weakly coupled to the lattice
phonons.\cite{Karvonen,KarvonenJP} Therefore, the electron-electron
scattering will cause nonequilibrium electrons to equilibrate on a
timescale that is typically much shorter than the electron-phonon
relaxation time, resulting in an electron distribution that is
thermal, but at a temperature higher than that of the lattice.
Understanding this hot-electron effect is crucial in studies of
transport in semiconductors and metals at low temperatures.\cite{Yu
and Cardona}

The electron-phonon scattering rate in conductors has been studied
both theoretically\cite{Little,Gantmakher review,Wellstood1994} and
experimentally,\cite{Wellstood1994,Roukes etal} with good agreement
between theory and experiment. The theoretical work has focused on
the rate of energy transfer between electrons and acoustic phonons
in an infinite bulk metal. The weak coupling between electrons and
phonons allows separate temperatures $T_{\rm el}$ and $T_{\rm ph}$
to be defined for the electron and phonon subsystems, and at low
temperatures significant differences in these temperatures are
easily achieved. In the deformation-potential couping, the primary
theoretical result states that the rate of energy exchange $P$ from
the electrons to the phonons is given by\cite{Wellstood1994}
\begin{equation}
P = \Sigma V \big(T_{\rm el}^n - T_{\rm ph}^n \big),
\label{Wellstood formula}
\end{equation}
where $n=5$, $V$ is the volume of the system, and $ \Sigma$ is a
material parameter given by
\begin{equation}
\Sigma= { 8\, \zeta(5) \, k_{\rm B}^5 \, \epsilon_{\rm F}^2 \,
N_{\rm el}(\epsilon_{\rm F}) \,  \over 3 \pi \hbar^4  \rho v_{\rm F}
v_{\rm l}^4} \label{Wellstood coefficient}
\end{equation}
Here $\zeta(m)$ is the Riemann zeta function, $\epsilon_{\rm F}$ is
the Fermi energy, $N_{\rm el}$ is the electronic density of states
(DOS) per unit volume, $\rho$ is the mass density, $v_{\rm l}$ is
the bulk longitudinal sound speed, and $v_{\rm F}$ is the Fermi
velocity.

All of the experiments to date, however, have been performed on thin
metal films deposited on semiconducting substrates, whereas the
theory has assumed bulk metals. The metal films typically have a
thickness $d$ of the order of ten to several hundred nanometers,
with lateral dimensions of a few microns. Although the electrons in
the films are three dimensional, the Fermi wavelength being much
smaller than any of the physical dimensions, the phonons at low
temperatures will have wavelengths larger than $d$. Therefore, one
expects similar temperature dependence of the thermal power
described by Eq.(\ref{Wellstood formula}), but with different
pre-factors $\Sigma$ and exponents $n$ ranging from $n=3$ to
$n=6$.\cite{Karvonen,KarvonenJP,DiTusaPRL92,Vinante2007} In
addition, there is a growing interest in the low-temperature
properties of electrons and phonons in mechanically suspended
nanostructures,\cite{Schwab
etal,Cleland2001,Cleland2002,Schmidt2004,Cross and
Lifshitz,Taskinen,Muhonen,Huard2007,Gusso} where the phonon spectrum
is strongly modified from that in a bulk metal, and thus the similar
behavior as in thin films was observed. Actually, experimental
results dictated that the reduction of phonon dimensionality leads
to weak temperature dependent of the heat flux between electrons and
phonons.\cite{Muhonen,Taskinen,Karvonen,DiTusaPRL92} In many cases,
the power transferred from electrons into phonons follows the $P\sim
T^{(D+2)}$ temperature dependence, where $D$ is the dimension of
space seen by the phonons.

%For Cu, the temperature dependence has been measured in the
%temperature range from 0.1 to 1K, and was found to be very close to
%that predicted by Eq.~(\ref{Wellstood formula}), and the prefactor
%has been measured to be $\Sigma_{\rm 3D} \approx 2 \times 10^9 \,
%{\rm W \, m^{-3} K^{-5}}$, in good agreement with the predicted
%value [given below in Eq.~(\ref{Wellstood coefficient}).]

The $T^3$ dependence of the heat flux obtained for suspended
one-dimensional metallic nanowires in Ref.~\onlinecite{Hekking2008}
also supports the above conjecture. It stimulates us to investigate
the hot-electron effect in a geometry of cylindrical nanoshell in
which electrons behave three-dimensionally but phonons are confined
in quasi-one dimension. This geometry can approximate a
single-walled metallic buckytube and also resemble the microtubules
found in many parts of the human body.\cite{Chapter7} Single-walled
carbon nanotubes are remarkable quasione-dimensional materials
because the nanotubes can be either metallic or semiconducting
depending on their helicity.\cite{Ishii} Therefore, applications of
single-walled carbon nanotubes as nanoscale electronic devices have
long been expected. In practice, metallic carbon nanotubes are very
good conductors and thus can be used to realize interconnection of
integrated circuits. Simple circuits based on semiconducting carbon
nanotube field-effect transistors have already been demonstrated.
The investigation of the electron-phonon interaction in nanotubes
plays an important role in understanding their properties and thus
has attracted much interest of
researchers.\cite{Ishii,Fouquet,Auer,Tsaousidou}. Hot carrier energy
relaxation is a promising candidate of probing the electron-phonon
coupling in carbon nanotubes because it is directly related to
inelastic scattering.\cite{Auer,Tsaousidou} In the current work, we
will derive an analytical expression for the net rate of thermal
energy, {\it i.e.} $P$, transferring from the electron to the phonon
subsystem in a cylindrical nanoshell in which the external pressure
on the inner surfaces is equal to that on the outer surfaces, where
the transverse dimensions are far smaller than the length in the
$z$-direction, and the thickness $h$ and the radius $R$ satisfy
condition $h \ll R$.

%However, most of the works are focused on the effects of coupling
%between electrons and optical phonons.
%little is done for the coupling between electrons and acoustic
%phonons because the contradiction in solving the phonon modes.

%The motion of the shell is directly related to many important
%processes from the point of physiology, such as cell movement and
%muscle contraction, and our interest is only about the
%electron-phonon coupling in it.

\section{general expression for the thermal power
\label{general expression}}

In Ref.~\onlinecite{Qu2005}, two of the authors have advanced a
general method for calculating the rate of thermal energy (thermal
power) transfer based upon a kind of weighted DOS. There, we
consider the hot electron effect in a system consisting of a
three-dimensional electron gas and a three-dimensional solid. The
Hamiltonian is
\begin{equation}
H = \sum_{\bf k} \epsilon_{\bf k} \, c_{\bf k}^\dagger c_{\bf k} +
\sum_n \hbar \omega_n \, a_n^\dagger a_n + \delta H,
\end{equation}
where $c_{\bf k}^\dagger$ and $c_{\bf k}$ are electron creation and
annihilation operators, with ${\bf k}$ the momentum, and
$a_n^\dagger$ and $a_n$ are bosonic phonon creation and annihilation
operators. The vibrational modes, labeled by an index $n$, are
eigenfunctions of the continuum elasticity equation
\begin{equation}
v_{\rm t}^2 {\bm \nabla} \times {\bm \nabla} \times {\bf u} - v_{\rm
l}^2 {\bm \nabla} ({\bm \nabla} \cdot {\bf u}) = \omega^2 {\bf u}
\end{equation}
for linear isotropic media, along with accompanying boundary
conditions. $v_{\rm t}$ and $v_{\rm l}$ are the bulk transverse and
longitudinal sound velocities. The electron-phonon interaction
$\delta H$ is described by the deformation potential
\begin{equation}
\delta H \equiv {{2 \over 3}} \epsilon_{\rm F} \! \int_{V_{\rm el}}
\! \! d^3r \, \psi^\dagger \psi \, {\bf \nabla} \cdot {\bf u},
\end{equation}
with
\begin{equation}
{\bf u}({\bf r}) = \sum_n (2 \rho \omega_n)^{-{1 \over 2}} [ {\bf
f}_n({\bf r}) \, a_n +  {\bf f}^*_n({\bf r}) \, a_n^\dagger ]
\end{equation}
the quantized displacement field. The vibrational eigenfunctions
${\bf f}_n({\bf r})$ are defined to be solutions of the elasticity
field equations, normalized over the phonon volume $V_{\rm ph}$
according to $\int_{V_{\rm ph}} \! \! d^3r \ {\bf f}_n^* \cdot {\bf
f}_{n'} = \delta_{nn'}.$ It will be convenient to rewrite the
electron-phonon interaction as
\begin{equation}
\delta H = \sum_{{\bf k q}n} [g_{n {\bf q}} \, c_{\bf k+q}^\dagger
c_{\bf k} \, a_n + g_{n {\bf q}}^* \, c_{\bf k-q}^\dagger c_{\bf k}
\, a_n^\dagger ],
\end{equation}
with coupling constant
\begin{equation}
g_{n {\bf q}} \equiv \frac{2}{3} \epsilon_{\rm F} (2 \rho
\omega_n)^{-{1\over2}} V_{\rm el} ^{-1} \! \int_{V_{\rm el}} \! \!
d^3r \ {\bf \nabla} \cdot {\bf f}_n \, e^{-i{\bf q} \cdot {\bf r}}.
\label{coupling constant}
\end{equation}

The quantity we calculate is the thermal energy per unit time
transferred from the electrons to the phonons,
\begin{equation}
P \equiv 2 \sum_{{\bf k q}n} \hbar \omega_n \big[ \Gamma^{\rm
em}_n({\bf k} \rightarrow {\bf k} - {\bf q}) - \Gamma^{\rm
ab}_n({\bf k} \rightarrow {\bf k} + {\bf q}) \big], \label{P
definition}
\end{equation}
where
\begin{eqnarray}
\Gamma^{\rm em}_n({\bf k} &\rightarrow& {\bf k} - {\bf q}) = 2 \pi
\, |g_{n {\bf q}}|^2 \, [n_{\rm B}(\omega_n) + 1]n_{\rm
F}(\epsilon_{\bf k})
\nonumber \\
&& \times[1-n_{\rm F}(\epsilon_{{\bf k}-{\bf q}}) ] \, \delta(
\epsilon_{{\bf k}-{\bf q}} - \epsilon_{\bf k} +  \omega_n )
\end{eqnarray}
is the golden-rule rate for an electron of momentum ${\bf k}$ to
scatter to ${\bf k} - {\bf q}$ while emitting a phonon $n$, and
\begin{eqnarray}
\Gamma^{\rm ab}_n({\bf k} &\rightarrow& {\bf k} + {\bf q}) = 2 \pi
\, |g_{n {\bf q}}|^2 \, n_{\rm B}(\omega_n)
n_{\rm F}(\epsilon_{\bf k}) \nonumber \\
&&\times[1-n_{\rm F}(\epsilon_{{\bf k}+{\bf q}}) ] \, \delta(
\epsilon_{{\bf k}+{\bf q}} - \epsilon_{\bf k} - \omega_n )
\end{eqnarray}
is the corresponding phonon absorption rate. $n_{\rm B}$ is the Bose
distribution function with temperature $T_{\rm ph}$ and $n_{\rm F}$
is the Fermi distribution with temperature $T_{\rm el}$. The factor
of 2 in (\ref{P definition}) accounts for spin degeneracy. It is
possible to obtain an exact expression for $P;$ the result
(suppressing factors of $\hbar$ and $k_{\rm B}$) is
\begin{eqnarray}
P&=&{m^2 V_{\rm el}^2 \over 8 \pi^4} \sum_{n}
\int_0^{\infty} d\omega\delta(\omega - \omega_n)
\big( {\textstyle{\omega \over e^{\omega/T_{\rm el}}-1}} -
{\textstyle{\omega \over e^{\omega/ T_{\rm ph}}-1}} \big) \nonumber \\
&\times&\int {\rm d^3}{\bf k}{|g_{n {\bf k}}|^2 \over |{\bf k}|} \bigg[ \omega +
T_{\rm el} \ln \bigg( {\textstyle{ 1 + \exp[({m \omega^2 \over 2
k^2} + {k^2 \over 8m} - {\omega \over 2} - \mu) / T_{\rm el}]) \over
1 + \exp[({m \omega^2 \over 2 k^2} + {k^2 \over 8m} + {\omega \over
2} - \mu) / T_{\rm el}]) }} \bigg) \bigg]. \nonumber
\label{general thermal power}
\end{eqnarray}
The logarithmic term in $P$ can be shown to be negligible in the
temperature regime of interest and will be dropped.

Although Eq.~(\ref{general thermal power}) is obtained for the
system consiting of 3-dimensional phonon and 3-dimensional electron
subsystems, we can simple extend it into those geometries that are
three-dimensional seen by electrons but arbitrary $D$-dimensional by
phonons. The DOS weighted form of $P$ (suppressing factors of
$\hbar$ and $k_{\rm B}$) reads
\begin{equation}
P \! = \! {2\,m^2 \epsilon_{\rm F}^2 \,\Omega_{\rm ph} \, \over 9 \,
\pi \rho} \int_0^{\omega_{\rm D}} \! \! \! \! \! d\omega \,
F(\omega) \bigg({\omega  \over e^{\omega/T_{\rm el}}-1} - {\omega
\over e^{\omega/T_{\rm ph}}-1} \bigg), \label{thermal power formula}
\end{equation}
where $\omega_{\rm D}$ is the Debye frequency, $\Omega_{\bf ph}$ is
the generalized volume of the $D$-dimensional phonon subsystem, and
$F(\omega)$ is a strain-weighted vibrational DOS, defined by
\begin{equation}
F(\omega) \equiv \sum_{n} U_n \, \delta(\omega - \omega_n),
\end{equation}
with
\begin{eqnarray}
U_n &\equiv& { 1 \over V_{\rm el}} \int_{V_{\rm el}} d^3r \, d^3r' \
{{\bm \nabla} \cdot {\bf f}_n({\bf r}) \ {\bm
\nabla}' \cdot {\bf f}_n^*({\bf r}')}\nonumber\\
&&{\hskip 0.20in}\times \int {d^D{\bf k}\over (2\pi)^D}\,{
k}^{-1}\,e^{-i{\bf k} \cdot ({\bf r}-{\bf r}^\prime)}.
\label{def_effective interaction}
\end{eqnarray}
Here $U_n$ can be interpreted as an energy associated with
mass-density fluctuations interacting via an potential determined by
the last integration in Eq.~(\ref{def_effective interaction}). To
this end we have reduced the calculation of $P$ to the calculation
of $F(\omega)$. Allen \cite{AllenPRL87} has derived a related
weighted-DOS formalism.

For the conventional hot-electron effect in bulk
materials,\cite{Wellstood1994} one deals with 3-dimensional phonon
subsystem and $\Omega_{\rm ph}=V_{\rm el}=V$. Caring out the last
integration in Eq.~(\ref{def_effective interaction}) gives
\begin{equation}
U_n = { 1 \over 2 \pi^2 V} \int_{V} \! d^3r \, d^3r' \ {{\bm \nabla}
\cdot {\bf f}_n({\bf r}) \ {\bm \nabla}' \cdot {\bf f}_n^*({\bf r}')
\over |{\bf r}-{\bf r}'|^2 }, \label{effective interaction_bulk}
\end{equation}
where the longitudinal acoustic phonon eigenfunction with momentum
${\bf k}$ is
\begin{equation}
{\bf f}_{\bf k} = {e^{i {\bf q} \cdot {\bf r}} \over \sqrt{V} } \,
{\bf e}_{\bf k}. \label{bulk phonon mode}
\end{equation}
Substituting this eigenfunction into Eq.~(\ref{effective
interaction_bulk}) and carrying out the integration, we obtain the
weighted DOS
\begin{equation}
F(\omega)={1\over 2 \pi^2 v_{\rm l}^4}\omega^3\label{bulk phonon
mode}
\end{equation}
The conventional expression for the net rate of energy relaxation in
Eq.~(\ref{Wellstood formula}) and the corresponding coefficient in
Eq.~(\ref{Wellstood coefficient}) can be easily reduced by this DOS.

\section{hot-electrons in a cylindrical nanoshell\label{cylindrical nanoshell}}
In this section we will derive the rate of the thermal energy
relaxation in a solid in which the electrons move
three-dimensionally but the phonons are confined in a cylindrical
nanoshell in which the external pressure on the inner surfaces is
equal to that on the outer surfaces, and the transverse dimensions
are far smaller than the length in the $z$-direction, and the
thickness $h$ and the radius $R$ satisfy condition $h \ll R$. There
has been extensively studies on their properties of the acoustic
phonons in cylindrical
shells.\cite{Chapter7,Sirenko1,Mahan2002,Sirenko2} In this work, we
employ the result of Ref.~\onlinecite{Chapter7}, in which the
eigenfunction reads,
\begin{eqnarray}
{\bf f}({\bf r})=
 \left(
  \begin{array}{ccc}
    -i C_r \\
    C_\varphi \\
    C_z \\
  \end{array}
\right)\exp(i m \varphi +iq_z z)
\end{eqnarray}
where $\varphi$ is the azimuth angular, $q_z\equiv q/R$ is the
wavevector in the $z$-direction. $C_r,C_\varphi$ and $C_z$ are
respectively the normalization constants in three orthogonal
directions. The dimensionless eigenmodes $\Omega=R\omega/c_l$ are
given by
\begin{eqnarray}
\Omega_m^{\rm I}(q) &\simeq& \sqrt{q^2+m^2+1},\\
\Omega_m^{\rm II}(q) &\simeq& \sqrt{\nu_{-}(q^2+m^2)},\\
\Omega_{m=0}^{\rm III}(q) &\simeq& \sqrt{1-\nu^2}{q\over{1+q}},\\
\Omega_m^{\rm III}(q) &\simeq&
\sqrt{1-\nu^2}{q^2\over{m\sqrt{m^2+1}+q^2}}
\end{eqnarray}
Here $q$ is the dimensionless wavevector, $m$ is the branch index,
$\nu_{-}=(1-\nu)/2$, $\nu$ is the Poisson ratio, and
$c_l=\sqrt{E/\rho(1-\nu^2)}$ is the longitudinal sound velocity (
$E$ is Young's modulus). In the axisymmetric case with $m=0$ the I,
II and III modes correspond to pure radial, torsional, and
longitudinal modes respectively. They correspond to pure
longitudinal modes, torsional, and radial respectively in the large
$q$ limitation.

Simply calculation via Eq.~(\ref{coupling constant}) reveals that
only the axisymmetric modes with $\Omega_0^{\rm I}$ and
$\Omega_0^{\rm III}$ contribute to the electron-phonon interaction.
The corresponding dispersion relations are
\begin{eqnarray}
\omega_{1}(q)&=& {c_l\over R}\Omega_0^{\rm I}=\omega_{\rm
c}\sqrt{q^2+1 }\label{Dispersion1} \\
\omega_{2}(q) &=& {c_l\over R}\Omega_0^{\rm III}=\omega_{\rm c}
{q\sqrt{1-\nu^2}\over {q+1}}\label{Dispersion2}
\end{eqnarray}
with $\omega_{\rm c}\equiv c_l/R$, the cutoff frequency for the
first mode. Here one may find that the minimum of $\omega_{1}$ is
always bigger than the maximum of $\omega_{2}$. The normalization
constants for the first mode are respectively $C_r=i\nu A_1 $,
$C_\varphi=0$ and $C_z=q A_1$ with $A_1\equiv[V_{\rm
el}(\nu^2+q^2)]^{-1/2}$. Those for the second mode are $C_r=-i \nu q
A_2$, $C_\varphi=0$ and $C_z=A_2(1+2q+\nu^2q^2)/(1+q)^2$ with
$A_2\equiv (1+q)^2 \{V_{\rm el}[\nu^2q^2(1+q)^4+(1+2q+\nu^2
q^2)^2]\}^{-1/2}$.

Obviously, phonons in this subsystem move quasi-one dimensionally
and thus we have $\Omega_{\rm ph}=L$ with $L$ the length of the
cylindrical nanoshell. Employing identity $ \int_{-\infty}^{\infty}
{{\rm d}{k}\, k^{-1}}\,e^{-i{k} (z-z^\prime)}=i \pi \,{\rm
sign}(z^\prime-z)$, Eq.~(\ref{def_effective interaction}) gives
\begin{eqnarray*}
U_{q_{z}}^m \equiv \frac{i}{2 V_{\rm el}}\int_{V_{\rm el}} d^3r \, d^3r' \ {{\bm \nabla} \cdot {\bf
f}_n({\bf r}) \ {\bm \nabla}' \cdot {\bf f}_n^*({\bf r}')}\,{\rm sign}(z^\prime-z),
\end{eqnarray*}
where the superscript $m={\rm 1}$ or $2$ distinguishes two
vibrational modes associated with the two dispersion relations
defined in Eq.~(\ref{Dispersion1}) and (\ref{Dispersion2}),
respectively. The strain-weighted vibration DOS is
\begin{eqnarray}
F(\omega)=F_1(\omega)+F_2(\omega)
\end{eqnarray}
%\begin{eqnarray}
%F(\omega)=\sum_{q_{z}}\, U_{q_{z}}^1\,\delta(\omega-\omega_1(q_z))
%+\sum_{q_{z}}\, U_{q_{z}}^2\,\delta(\omega-\omega_2(q_z)).\nonumber
%\end{eqnarray}
with $F_m(\omega)$ defined by
\begin{eqnarray}
F_m(\omega) &\equiv& \sum_{q_{z}}\,
U_{q_{z}}^m\,\delta(\omega-\omega_1(q_z))\nonumber \\
&=&{ i L \over 4 \pi R V_{\rm el}} \int_{-\infty}^{\infty} {\rm d}
q \delta(\omega-\omega_m(q)) \nonumber \\
{\hskip -0.20in}&\times& {\hskip -0.10in}\int_{V_{\rm el}} d^3 {\bf
r} \, d^3 {\bf r}' \ {{\bm \nabla} \cdot {\bf f}_n({\bf r}) \ {\bm
\nabla}' \cdot {\bf f}_n^*({\bf r}')}\,{\rm
sign}(z^\prime-z).\nonumber
\\ \label{Fm}
\end{eqnarray}
Here, $F_1(\omega)$ is valid in the range confined by $\omega \geq
\omega_{\rm c}$. Carrying out the integrations over ${\bf r}$, ${\bf
r}'$, and $q$, we obatin
\begin{equation}
F_1(\omega) = \left({\omega\over2 \pi c_l^2}\right) { 1-{\omega^2/
\omega_{\rm c}^2 } \over {(1-\nu^2)-\omega^2/ \omega_{\rm c}^2}}
\label{DOS_F1}
\end{equation}
and
\begin{eqnarray}
F_2(\omega) &=& \left({\omega\over 2 \pi c_l^2}\right) {
\sqrt{1-\nu^2}\over {\sqrt{1-\nu^2} - \omega/ \omega_{\rm c}}}
\nonumber\\
&& {\hskip -0.20in}{\left(1-\omega^2/ \omega_{\rm c}^2\right)^2
\over {\nu^2 \omega^2 / \omega_{\rm c}^2
 + \left(1-\omega^2 / \omega_{\rm c}^2\right)^2
\left(\sqrt{1-\nu^2}-\omega / \omega_{\rm c}
\right)^2}}.\nonumber\\\label{DOS_F2}
\end{eqnarray}
Then the thermal power can be calculated by the following formula,
\begin{eqnarray}
P&=&\frac{2m^2 \epsilon_{F}^2 \,L }{9 \pi\rho}\bigg[\int_{\omega_{c}}^{\omega_{D}}
d\omega F_1(\omega)\bigg({\omega  \over e^{\omega/T_{\rm
el}}-1} -{\omega \over e^{\omega/T_{\rm ph}}-1} \bigg)\nonumber\\
&+&\int_{0}^{\omega_{D}}d\omega F_2(\omega)\bigg({\omega \over e^{\omega/T_{\rm el}}-1}
-{\omega \over e^{\omega/T_{\rm ph}}-1} \bigg)\bigg].
\label{thermal power formula_shell}
\end{eqnarray}

 The strain-weighted DOS has distinct high- and low-frequency character
crossing over at $\omega=\omega_{\rm c}$ because $F_1(\omega)$ has a
cut off at this frequency. Clearly, $\omega_{\rm c}$ defines a
nature crossover temperature
\begin{equation}
T^\star \equiv {\hbar\,  \omega_{\rm c} \over {k_{\rm B} }}
\end{equation}
distinguishing the high and low temperature behaviors.

%\subsection{High-Temperature Limitation}
At high temperature, $F_1({\omega})$ dominates the contribution, and
the large frequency asymptotic behavior of strain-weighted DOS reads
\begin{equation}
F(\omega)=F_1(\omega)\simeq{\omega\over{2\pi c_l^2}},\nonumber
\end{equation}
leading to the high-temperature expression of thermal power
\begin{equation}
P = \Sigma_{\rm 1D} \, L\big[ \Phi_3(\omega_{\rm D}/T_{\rm el}) \,
T_{\rm el}^3 - \Phi_3(\omega_{\rm D}/T_{\rm ph}) \, T_{\rm ph}^3
\big],\label{high temperature form of P}
\end{equation}
here $\Sigma_{\rm 1D}$ is the material parameter, defined by
\begin{equation}
\Sigma_{\rm 1D}={\zeta(3) k_{\rm B}^3 k_{\rm F}^4\over{18\rho \hbar
\pi^2 c_l^2}},
\end{equation}
with $\Phi_m(y) \equiv [(m-1)! \, \zeta(m)]^{-1} \! \int_0^y dx \
x^{m-1} /(e^x - 1)$. It can be shown that the $\Phi_3$ factors are
equal to unity at temperatures above $T^\star$ but sufficiently
smaller than the Debye temperature. Thus, expression~(\ref{high
temperature form of P}) reduces to
\begin{equation}
P = \Sigma_{\rm 1D} \, L \left(T_{\rm el}^3 - T_{\rm ph}^3
\right).\label{very high temperature form of P}
\end{equation}
The $T^3$ law for the thermal power for cylindrical nanoshells
revealed by this relation is the same to that in one dimensional
phonon systems where electron cooling devices are
manipulated.\cite{Hekking2008,Muhonen} The crossover from the $T^5$
law in three-dimensional phonon systems to the $T^3$ law in their
one-dimensional counterparts shows that the electron-phonon
interaction changes very significantly when dimensionality of the
phonon subsystem reduces, which agrees with the effect of phonon
dimensionality on the electron-phonon coupling observed by other
researches.\cite{Hekking2008,Karvonen,Muhonen,Vinante2007}

Now we apply our theory to a metallic carbon nanotube with the
Poisson ratio\cite{Liu2007} $\nu=0.186$, the Young's
modulus\cite{Liu2007} $E=1050\,{\rm Gpa}$, the mass
density\cite{Mahan2002} $\rho=2.26\,{\rm g/cm^3}$, and the Fermi
energy\cite{Lin1993} $\epsilon_{\rm F}=1.0 \,{\rm eV}$. The
calculated longitudinal sound velocity is $c_l=21.9\times10^3\,{\rm
m/s}$ and the the material parameter $\Sigma_{\rm
1D}=1.08\times10^{-10}\,{\rm Wm^{-1}K^{-3}}$.

%\subsection{Low-Temperature Limitation}
At low-temperature when $T < T^\star$, the dominant contributions to
the thermal power are due to the small frequency asymptotical
behavior of $F(\omega)$, which are
\begin{eqnarray}
F_1(\omega) &\simeq& {\omega\over{2\pi c_l^2}},\,\,\,\,\omega \ge \omega_{\rm c}\nonumber\\
F_2(\omega) &\simeq& {\omega \over {2\pi c_l^2}}\left[{1\over
{1-\nu^2}}+{3\omega\over {\omega_{\rm
c}(1-\nu^2)^{3/2}}}\right].\nonumber
\end{eqnarray}
Combining these two expressions with Eq.~(\ref{thermal power
formula_shell}), we get the low-temperature expression of the
thermal power, {\it i.e.},
\begin{eqnarray}
P&=&{\Sigma_{\rm 1D} L} \left[ \Lambda(T_{\rm el})\,T_{\rm
el}^3-\Lambda(T_{\rm ph})\,T_{\rm ph}^3\right]
\nonumber\\
&&{\hskip -0.15in}+{\Sigma_{\rm 1D} L}\left[\Pi(T_{\rm el})\,T_{\rm
el}^4-\Pi (T_{\rm ph})\,T_{\rm ph}^4\right]\label{low temperature
form of P}
\end{eqnarray}
where
\begin{eqnarray}
\Lambda(T) &\equiv& {2\Phi_3({\omega_{\rm D}}/T)-\Phi_3({\omega_{\rm
c}}/T) \over {1-\nu^2}}, \nonumber \\
\Pi(T) &\equiv& {9\gamma \over {T^\star (1-\nu^2)^{3/2}}}\Phi_4
({\omega_{\rm D}}/T),
\nonumber \\
\gamma &\equiv& \zeta(4)/\zeta(3).\nonumber
\end{eqnarray}

At extremely low temperature, {\it i.e.} $T\ll T^\star$, $\omega_D/T
\gg 1$, and thus $\Phi_3$ and $\Phi_4$ factors tend to unity.
Therefore, we have $\Lambda(T)=1/(1-\nu^2)$ and
$\Pi(T)=9\gamma/T^\star (1-\nu^2)^{3/2}$, and hence
expression~(\ref{low temperature form of P}) reduces to
\begin{equation}
P=\Sigma_{\rm 1D} L \left[{T_{\rm el}^3 - T_{\rm ph}^3\over
{1-\nu^2}} +{9 \gamma \left(T_{\rm el}^4-T_{\rm ph}^4\right) \over
{T^\star (1-\nu^2)^{3/2}}}\right] \label{very low temperature form
of P},
\end{equation}
which reveals a crossover of the $T^3$ law for the temperature
dependence of the thermal power at low temperature. The crossover
temperature for the metallic carbon nanotube is $T^\star =24.6 \,K$,
which is in the temperature zone that the current technology can
achieve. Hence we expect the experimental observation of this
crossover phenomenon.

%\begin{figure}
%\includegraphics[width=6.5cm]{Graph1.eps}
%\caption{\label{Gth-R1.001-b100nm figure} $\Phi(y)$ function.}
%\end{figure}
%\begin{figure}
%\includegraphics[width=6.5cm]{Graph2.eps}
%\caption{\label{Gth-R1.001-b100nm figure} the relationship between
%$P$ and $T$.}
%\end{figure}

\section{Conclusion}
We have advanced a general expression for calculating the thermal
power transferring from 3-dimensional electron to any
$D$-dimensional acoustic phonon subsystem. It transfers the
calculation of thermal power to the computation of the
strain-weighted DOS for phonon subsystems of any dimensionality.
Employing this method, we have investigated the hot-electron effect
in free suspended cylindrical nanoshells, in which electrons behave
three-dimensionally but phonons are confined to quasi-one dimension.
The temperature dependence of the thermal power is obtained
analytically, and the low-temperature crossover from the $T^3$ to
$T^3/(1-\nu^2)+9\gamma T^4/[T^*(1-\nu^2)^{3/2}]$ dependence is also
deduced. The result shows that reduction of phonon dimensionality
leads to weak temperature dependence of the heat flux between
electrons and acoustic phonons. Even though we only estimate the
corresponding quantities for the material parameters from a metallic
carbon nanotube, the formula developed in this work is also
applicable to any kind of the cylindrical shells in nanoscale, where
the coupling between electrons and acoustic phonons is important.

\acknowledgments {This work was supported by the Key Project of
Chinese Ministry of Education.(No.108118)}

\end{document}